\documentclass[prl,twocolumn,amsmath,amssymb,floatfix,showpacs]{revtex4}
\usepackage{graphicx}
\usepackage{bm}
\usepackage{amssymb}
\usepackage{color}
\usepackage{epsfig}
\usepackage{subfigure}

\newcommand{\be}{\begin{equation}}
\newcommand{\ee}{\end{equation}}
\begin{document}

\title {A new universality class of aging in glassy systems}

\author{Ariel Amir, Yuval Oreg, Yoseph Imry}

\affiliation { Department of Condensed Matter Physics, Weizmann
Institute of Science, Rehovot, 76100, Israel\\}

\begin{abstract}
Glassy systems are ubiquitous in nature, and are characterized by
slow relaxations to equilibrium without a typical timescale, aging
and memory effects. Understanding these is a long-standing problem
in physics. We study the aging of the electron glass, a system
showing remarkable slow relaxations in the conductance. We find that
the broad distribution of relaxation rates leads to a universal
relaxation of the form $\log(1+t_w/t)$ for the usual aging protocol,
where $t_w$ is the length of time the perturbation driving the
system out-of-equilibrium was on, and $t$ the time of measurement.
These results agree well with several experiments performed on
different glasses, and examining different physical observables, for
times ranging from seconds to several hours. The suggested
theoretical framework appears to offer a paradigm for aging in a
broad class of glassy materials.

\end{abstract}


 \maketitle

Aging is one of the most distinct characteristics of glasses.
Crudely speaking, it is the dependence of the form of the relaxation
to equilibrium on the time an external perturbation was acting on
the system. Aging and slow relaxations have been experimentally
observed in various systems. These range from
spin-glass~\cite{spinglass}, to structural glasses~\cite{structural,
structural_ludwig}, gels~\cite {gels}, vortices in
superconductors~\cite {superconductorglass},
ferroelectrics~\cite{ferroelectric}, colloids and granular systems
~\cite{granular} and the electron-glass system~\cite{zvi}. Analytic
models and simulations showing aging have been of considerate
interest in the past decades~\cite{aging_models}. In general, the
return to equilibrium can be characterized by a function $F(t,t_w)$,
where $t_w$, the 'waiting-time', is the time the perturbation was on
(during which the system is being 'aged'), and $t$ is the time from
switching off the perturbation until the measurement. We refer to
the complete protocol as an 'aging experiment'. Full (or simple)
aging is said to hold when $F(t,t_w)=f(t_w/t)$.

 Experiments performed on
the electron glass system, measuring the conductance as a function
of time when the system is perturbed, show remarkable data collapse
when time is measured in units of $t_w$~\cite{zvi}. This proves
experimentally that full aging is obeyed. For times $t \ll t_w$ a
logarithmic relaxation is observed. At times $t$ of order $t_w$
deviations from the logarithm are seen. Similar results were
obtained for granular aluminum~\cite{Grenet} and structural
glasses~\cite{structural_ludwig} which suggest that the physical
mechanisms leading to such logarithmic relaxations may be
ubiquitous. In this paper we suggest a model which should be
applicable for several types of physical realizations, and yields
the result that full aging indeed holds. Moreover, it gives under
certain simplifying assumptions $f(x) \sim \log(1+x)$, which fits
the experimental data well for several decades in time.

\emph{The Model}.- To start with, let us consider $N$ localized
disordered states and $M < N$ interacting electrons, with a coupling
between the electrons and a phonon reservoir~\cite{amir_glass}.
After ensemble averaging over the phonon reservoir, the averaged
occupation numbers can take any value between 0 and 1, and their
dynamics is described by a set of rate~equations: $\frac{dn_i}{dt} =
\sum_{j \neq i} \gamma_{j,i}- \gamma_{i,j} , \label {dynamics} $
where due to the Coulomb interaction and the Pauli principle the
rates $\gamma_{i,j}$, which describe the phonon-assisted transitions
between electronic states, depend on the set of occupation numbers
$\{n_i\}$. Linearization close to (one of the many) metastable
states leads to an equation $\frac{d\vec{\delta n}}{dt}=A
\vec{\delta n}$, where $\vec{\delta n}$ is a vector of the
deviations from the metastable state, and $A$ a matrix depending on
the properties of the metastable state. Ref.
[\onlinecite{amir_glass}] explains in detail the procedure, and
shows that the distribution of eigenvalues of the matrix is given
approximately by $P(\lambda) \sim 1/\lambda$ within a large
frequency window, essentially arising from the exponential
dependence of the matrix elements on the physical parameters
(distance and energy difference). The $1/\lambda$ distribution leads
to a logarithmic decay in time, as can be seen by Laplace
transforming it.

From now on we consider quite generally a system whose dynamics is
described by a linear relation $\frac{d\vec{\delta n}}{dt}=A \cdot
\vec{\delta n}$, and with the eigenvalue distribution of the matrix
$A$ described approximately by $P(\lambda) \sim 1/\lambda$. The
mean-field approximation leads to this result starting from a
microscopic Hamiltonian, but we believe that the results should
apply for a broader range of models. At the end of the paper we
demonstrate this by applying the theory to the experimental data of
Ref. [\onlinecite{structural_ludwig}].

We will assume that the system contains another external control
parameter we shall call $V_g$, such as the gate voltage in the
electron glass experiment~\cite{zvi}. Changing an external system
parameter will shift the metastable states' positions. Thus, upon
small changes in $V_g$, the linearized matrix $A$ will slightly
change. Let us consider the system in two metastable states $a$ and
$b$ corresponding to values $V_a$ and $V_b$ respectively. We shall
denote the linearized matrix $A$ in the state $a$($b$) by $A^a$
($A^b$), and the eigenvectors $\vec{a}_q$ (${\vec{b}_q}$). The
metastable state $a$ is characterized by a configuration
$\vec{n}_a$, and the state $b$ by a configuration
$\vec{n}_b=\vec{n}_a+\Delta \vec{n}$.

The experimental protocol is as follows:

I The gate voltage is $V_a$. The system is assumed to 'equilibrate'
to a metastable state.

II The gate voltage is changed to $V_b=V_a+ \delta V$, the system
'ages' for a time $t_w$: the system approaches the new equilibrium
$b$.

III The gate voltage is changed back to a value $V_a$. The system is
in the process of returning to the original equilibrium $a$.

The conductance is measured throughout the experiment. The
measurement time is much shorter than the relevant relaxation times.
For convenience we define $t=0$ at the end of stage II, when the
gate voltage is changed to its initial value $V_a$.

\begin{figure}[h]
\begin{center}
$\begin{array}{c@{\hspace{0.00in}}c} \multicolumn{1}{l} {\mbox{}} &
\multicolumn{1}{l}{\mbox{ }} \\ [-0.0cm] \epsfxsize=3.5in
\hspace{0.26 in}\epsffile{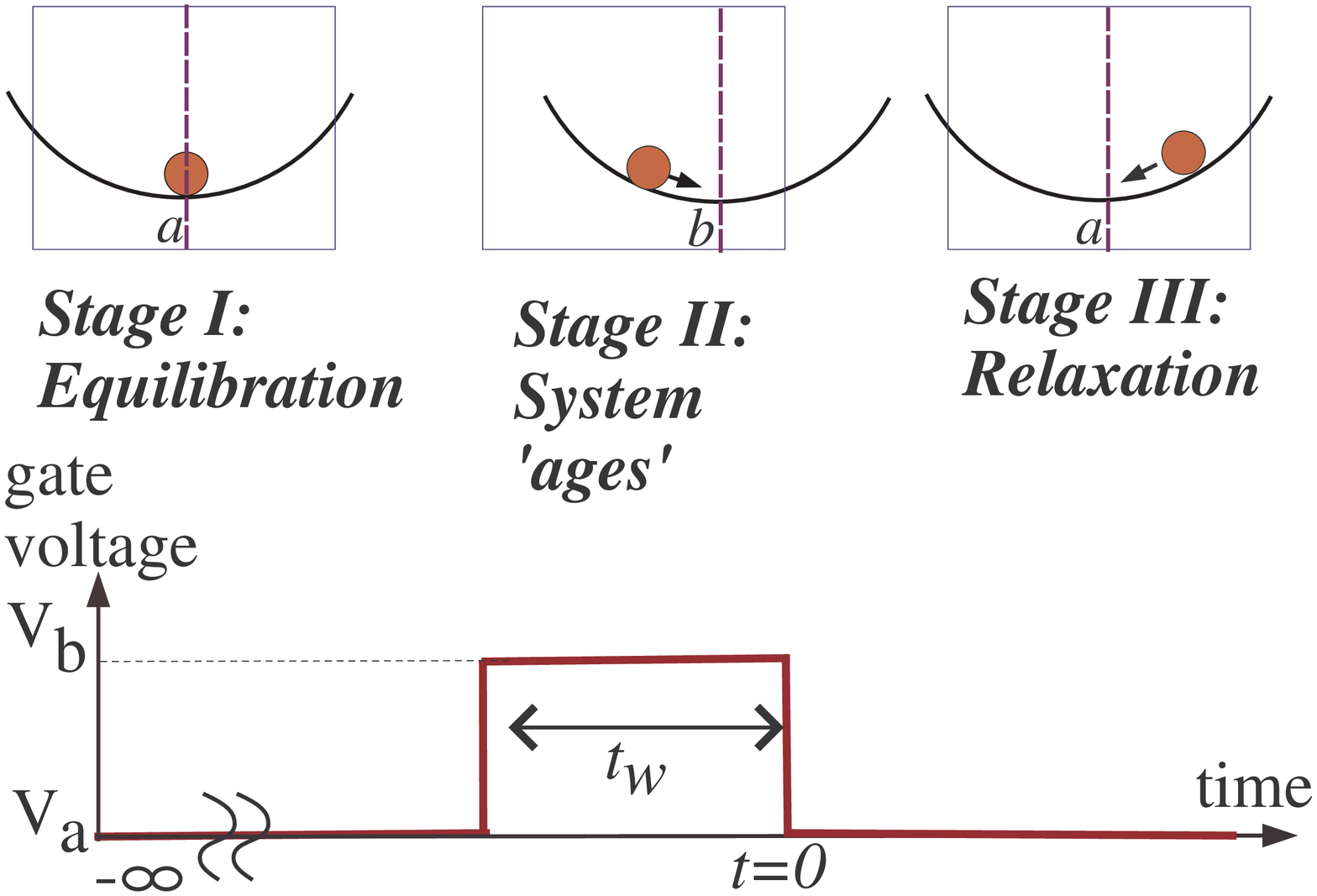} \\
\hspace{-0.7 in}\epsfxsize=3.6in \epsffile{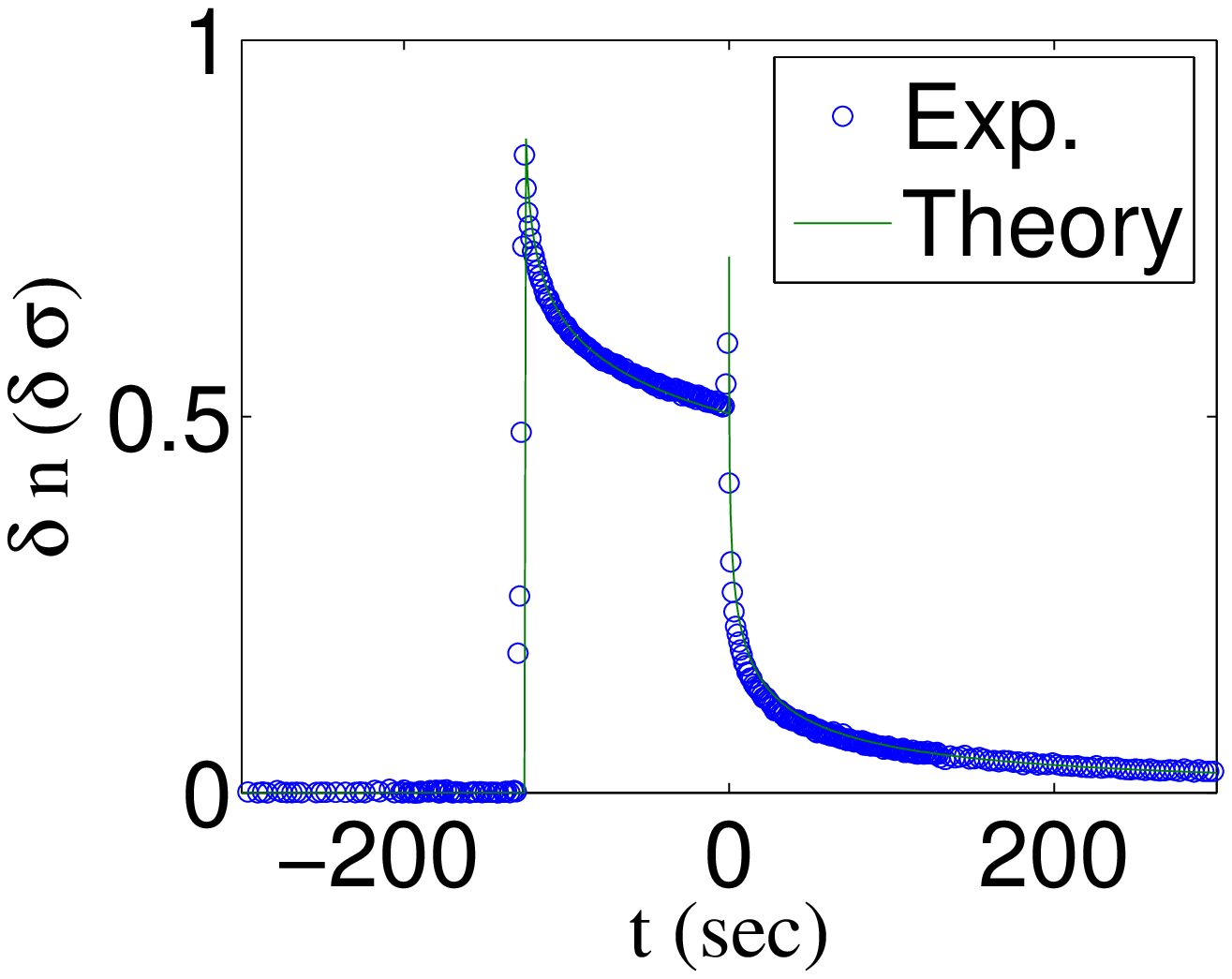} \\
\end{array}$
\end{center}

\vspace{0 cm} \caption{ Schematic demonstration of the aging
process. Initially, the system is in a metastable configuration $a$.
When the landscape changes due to a change of gate voltage, the
system relaxes towards the new minimum $b$, during time $t_w$. When
the landscape returns to its initial form, the system relaxes to
$a$. Below we schematically show our results, for the time
dependence of the distance to the current metastable state ($a$ or
$b$) $\delta n (t)$. At first, the system sits at a metastable
state. At a certain time (-200 sec. for the figure) the system
parameter (i.e., gate voltage) is changed, and $\delta n(t)$ jumps
and then relaxes logarithmically. At time $t_w$ the system parameter
is changed to its initial value, from which the relaxation takes the
form $\log(1+t_w/t)$ (Eq. (\ref{log_mod})). The inset shows the
experimental result for a typical electron glass aging experiment
(courtesy of Z. Ovadyahu). The lower cutoff used in the theory,
relevant only for stage II, is a fitting parameter, and was taken to
be 50 hours. \label{aging_fig} }
\end{figure}

During stage I, the system reaches, presumably, one of the many
metastable states ($a$), and lingers there. At the beginning of
stage II ($t=-t_w$) we change the potential landscape by shifting
suddenly the gate voltage from $V_a$ to $V_b$. The occupations at
this time are still equal to the metastable state occupations
$\vec{n}_a$. \emph{During} stage II it starts, however, to relax to
the new metastable state $b$. The new metastable state $b$ is not
far (in phase space) from the original one $a$ , but since the
relaxation to it is composed of a broad range of timescales, the
system takes a long time to fully equilibrate to state $b$. At the
beginning of stage III ($t=0$) the gate voltage is changed back to
its initial value. During stage III, the system is relaxing back to
$a$, the metastable state it was initially in~\cite{clarification}.
This process is described schematically in Fig. \ref{aging_fig},
along with our theoretical predictions.

Let us proceed to the calculation. In stage I, the system is in the
metastable configuration $\vec{n}_a$. At the beginning of stage II,
the system is in the state $\vec{n}_a=\vec{n}_b-\Delta \vec{n}$.
Working in the eigenbasis of $b$, $-\Delta \vec{n} = \sum_q c_q
{\vec{b}_q} . \label{cq} $ This equation determines the
decomposition weights $c_q$ (which are not simply the projections
onto the eigenvectors, since the matrix $A^b$ is not
hermitian~\cite{amir_glass}). Therefore during stage II (the aging
process), the configuration of the system is described by:

\be \vec{n}(t)=\vec{n}_b+ \sum_q c_q {\vec{b}_q} e^{-\lambda^b_q
(t+t_w)}, -t_w<t<0, \ee where $\lambda^b_q$ are the relaxation rates
(the eigenvalues of the linearized matrices $A^a$ and $A^b$ are real
and negative, corresponding to pure decay). Since the eigenvectors
are localized~\cite{amir_glass}, we can assume that $c_q$ are random
variables whose distributions and typical values do not depend
explicitly on $q$. Thus, for a large system: $|\sum_q c_q
{\vec{b}_q}e^{-\lambda^b_q \Delta t}| \sim
\int_{\lambda_{min}}^{\lambda_{max}} \frac{e^{- \lambda \Delta t}
}{\lambda} d \lambda = Ei (\lambda_{max} \Delta
t)-E_i(\lambda_{min}\Delta t) ,$ where $Ei(x)$ is the exponential
integral function. For times $\Delta t\rightarrow 0$ the integral
approaches a constant, which by construction is $|\Delta \vec{n}|$,
while for times $\Delta t\gg 1/\lambda_{min}$ it falls off
exponentially. The interesting behavior occurs for the large time
window $1/\lambda_{max}<\Delta t < 1/\lambda_{min}$, where we can
approximate the integral by $\gamma_E-\log [ \Delta t \lambda_{min}]
,\label{logarithm}$ with $\gamma_E$ is the Euler constant. This
implies that during stage II we should observe a logarithmic
relaxation, which is indeed observed experimentally~\cite{zvi:4}.

 In stage III, it is convenient to work in the eigenvector basis of metastable state $a$. We can
express the configuration of the system at the end of stage II as
$\vec{n}(t=0)=\vec{n}_a+ \Delta \vec{n}+ \sum_q c_q {\vec{b}_q}
e^{-\lambda^b_q t_w}.$ Since the perturbation is small, to lowest
order we can replace ${\vec{b}_q}$ by ${\vec{a}_q}$ and substitute
the eigenvector decomposition of $\Delta n$, obtaining: $
\vec{n}(0)=\vec{n}_a+\sum_q c_q {\vec{a}_q}(e^{-\lambda^b_q t_w}-1).
\label{stretched_modes}$ Note that during the waiting-time, each of
the eigenmodes approaches $b$ with a characteristic time
$1/\lambda_q$. Therefore only modes obeying $\lambda t_w \gtrsim 1$
had enough time to relax to $b$. When the voltage is changed to its
initial value, this relaxation to $b$ is in fact an
\emph{excitation} with respect to the metastable state $a$. Thus,
slowly relaxing modes are also slow to be excited, which is the crux
of the matter. The relaxation back to equilibrium during the last
stage of the protocol is described as: $ \vec{n}(t)=\vec{n}_a+\sum_q
c_q {\vec{a}_q}(e^{-\lambda^b_q t_w}-1)e^{-\lambda^a_q t}.$

We can approximate $\lambda^a_q \approx \lambda^b_q \equiv
\lambda_q$, and obtain for the occupancy vector at time $t$: $
\delta \vec{n}(t)=\sum_q c_q {\vec{a}_q}(e^{-\lambda_q
t_w}-1)e^{-\lambda_q t}.$ Defining $ \delta n(t) \equiv |\delta
\vec{n}(t)| $, we obtain:

\be \delta n(t) =C[\gamma_E- \log(t+t_w)] - C[\gamma_E-\log(t)] = C
\log(1+t_w/t), \label{log_mod}\ee where $C=\frac{|\Delta
\vec{n}|}{\log(\lambda_{max}/\lambda_{min})}$ is a non-universal
constant, which depends on the system parameters.

Notice that $|\delta \vec{n}(t)|$ is a function of $t/t_w$,
corresponding to full aging. This is due to the fact that for $f(x)
\sim \log(x)$, the difference $f(t+t_w)-f(t)$ is only a function of
the ratio ${t}/{t_w}$. It can be proven that \emph{only} a
logarithmic function has this characteristic. Since $f(x)$ was
essentially a Laplace transform of the eigenvalue distribution, this
shows a remarkable relation between the ${1}/{\lambda}$ distribution
and full aging.
%
Fig. \ref{aging_fig} summarizes the theoretically expected time
dependence throughout the experiment, and shows the results of a
typical electron glass aging experiment.

So far we have discussed the aging of the average occupation
numbers. To relate this to the measured observable in the electron
glass experiments, the conductance, we need to know the relation
$\delta \sigma(\delta n)$. We emphasize that for any such relation
we would obtain full aging, since $\delta n$ has full aging. In the
simplest approximation, one would assume a linear relation between
the two. In other words $\delta n(t)$, essentially the distance in
phase space to the local minimum (the metastable state), is
proportional to the excess conductivity. This may arise due to
various mechanisms. One may intuitively regard perturbation in
occupation number as an effective excess temperature, raising the
conductivity. A more subtle mechanism would arise due to slowly
relaxing modes, influencing the conductivity of the system via the
long-range Coulomb interactions~\cite{{yu},{amir_glass}}. This is an
important point, that may distinguish the electron glass system from
spin glasses,
and we intend to elaborate on it in future work. 

\begin{figure}[b!]
\includegraphics[width=0.35\textwidth]{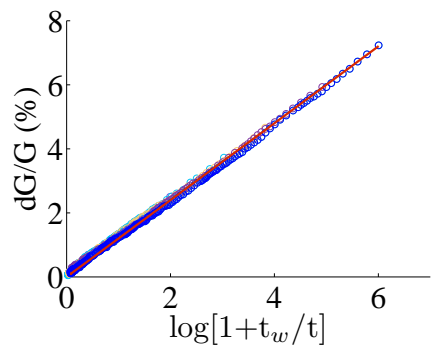}
\caption{ The excess conductance is plotted as a function of the
theoretical expression $\log (1+{t_w}/{t})$ . There are no fitting
parameters in this expression. The data shown is for 5 experiments,
corresponding to different colors, with different waiting times
$t_w$ of 20, 60, 180, 540 and 1620 seconds. The ratio ${t}/{tw}$
varies from 0.0025 to 36. We thank Z. Ovadyahu for providing us with
the data. \label{aging} }
\end{figure}

 \emph{Comparison to experiment}.- Fig. \ref {aging}
compares the theoretical form in stage III and the experimental data
of Ovadyahu's group \cite{zvi}. The experimental data shows a
logarithmic regime for times shorter than $t_w$, and power-law
regime, for times much longer than $t_w$. These two limits, as well
as the crossover between them, are contained in the theoretical
expression. The significance of the fit shown in the figure is the
linear dependence of the excess conductivity on the predicted
function $\log(1+{t_w}/{t})$. There are no fitting parameters other
than the proportionality constant, which is non-universal.

The logarithmic behavior predicted for short times is an important
distinction between the theory we present and various theories
predicting a stretched exponential behavior,
$e^{-\left(\frac{t}{t_0}\right)^\alpha}$~\cite{stretched_exponential}.
At short times, compared to $t_w$, the stretched exponential reduces
to a power-law, which is inconsistent with the experimental data, as
was demonstrated in [\onlinecite{zvi:4}].

For even longer times (on the scale set by $t_w$), deviations from
the theory occur. These can be corrected by taking a non-linear
relation between $\delta n$ and $\delta \sigma$. In order to fit
correctly the experimental data, the corrected relation must be
sublinear. An additional sublinear term proportional to
$\sqrt{\delta n}$ provides excellent correspondence to the
experimental data (not shown). That such a sublinear relation may
exist is supported by a different experiment, where the conductance
is measured as a function of the gate voltage in a relatively fast
scan~\cite{{zvi:2},{muller}}.


%

\emph{Generality of the theory}.- 
It is possible that this model is in fact only a caricature of the
real scenario: possibly, the system explores many metastable states
(local minima of the potential landscape). If we consider the vector
of the probabilities to remain in each state, the transition between
different minima are also exponentials of a broadly distributed
quantity~\cite{kogan}. This implies that the 'ingredients' used in
our calculation are still valid: if we consider the vector of
probabilities of the system to be in each of the minima, the
(classical) dynamics of the system will be described by a master
equation of the type $\frac{d\vec{P}}{dt}=A \vec{P}$, where
$\vec{P}$ is a vector of probabilities to be at different minima. A
wide, slowly varying, distribution of barriers would again lead to a
${1}/{\lambda}$ distribution of eigenvalues of the matrix $A$,
leading to the relaxation described by Eq. (\ref{log_mod}).

We demonstrate the generality of the theory by analyzing a different
experiment. Ludwig \emph{et al.} \cite{structural_ludwig} measured
the dielectric constant of Mylar when pushed out-of-equilibrium,
using a similar aging protocol to the one described. We consider the
data from the first measurement in each of their aging experiments,
for which the system is fully equilibrated. The data presented in
the first reference of [\onlinecite{structural_ludwig}] shows that
for times short compared to $t_w$, the signal is proportional to
$\log(c{t_w}/{t})$, with $c=1.02$ extracted from the data. This fits
well with our prediction of $\log(1+{t_w}/{t})$. We emphasize that
in contrast to our theory, the theory used in Ref.~\cite{burin}
gives a non-universal value of $c$, depending on the system
parameters.

\emph{Summary}.- We have shown how full aging may naturally occur in
a model for the electron glass, starting with a microscopic picture,
and using a master equation to analyze the dynamics of the system.
Under simplifying assumptions, we have shown that the deviation from
the metastable state should follow a universal form $\sim \log(1+
{t_w}/{t})$, where $t_w$ is the length of time the perturbation
driving the system out-of-equilibrium was on, and $t$ the time of
measurement. This agrees well with experimental data of the
conductance relaxation in the electron glass InO, within a large
time window, ranging from seconds to several hours. The theoretical
predictions also explain experiments performed on a totally
different glass, the plastic Mylar, where the dielectric constant is
measured. Although the system is completely different, the
universality of the underlying distribution of relaxation rates
$P(\lambda) \sim {1}/{\lambda}$ lead to our general result for the
relaxation. This suggests that our model of relaxation arising from
a broad ${1}/{\lambda}$ spectrum may be a paradigm for aging in
various glasses.

We thank Z. Ovadyahu and S. Ludwig for kindly providing us with
their experimental data, as well as for illuminating discussions.

This work was supported by a BMBF DIP grant as well as by ISF and
BSF grants and the Center of Excellence Program. A.A. acknowledges
funding by the Israel Ministry of Science and Technology via the
Eshkol~program.


\begin{thebibliography}{10}
\bibitem{clarification}
We do not claim that the system will return to the same microscopic
  configuration: the mean-field approximation captures the ensemble average,
  and thus characterizes the statistical average. Since we are interested in
  the conductivity of the system, which is a macroscopic property, it is
  sufficient to understand the ensemble-averaged behavior.
\end{thebibliography}
\end{document}